\documentclass[conference]{IEEEtran}
\IEEEoverridecommandlockouts
\usepackage{cite}
\usepackage{amsmath,amssymb,amsfonts}
\usepackage{algorithmic}
\usepackage{graphicx}
\usepackage{textcomp}
\usepackage{xcolor}
\usepackage{bm}
\usepackage{hyperref}
\usepackage{threeparttable}
\usepackage{cite}
\usepackage{makecell}
\usepackage{hyperref} 
\usepackage{amssymb}
\usepackage{amsmath}
\usepackage{amsthm}
\usepackage{bm}
\usepackage{float}
\usepackage{color}
\usepackage{colortbl}
\usepackage{scalerel,stackengine}
\usepackage{multicol}
\usepackage{multirow}
\usepackage{graphicx}
\usepackage{diagbox}
\usepackage{multirow}
\usepackage{tabularx}

\usepackage{balance}

\def\BibTeX{{\rm B\kern-.05em{\sc i\kern-.025em b}\kern-.08em
    T\kern-.1667em\lower.7ex\hbox{E}\kern-.125emX}}

\title{Efficient Probabilistic Optimal Power Flow Assessment Using an Adaptive Stochastic Spectral Embedding Surrogate Model

\thanks{This work was supported partially by the Natural Sciences and Engineering Research Council (NSERC) Discovery Grant, NSERC RGPIN-2022-03236 and partially by the Fonds de Recherche du Qu\'{e}bec-Nature et technologies under Grant FRQ-NT 320645.}
}

\date{September 2023}

\begin{document}
\author{Xiaoting Wang,~\IEEEmembership{Student Member,~IEEE}, Jingyu Liu,~\IEEEmembership{Student Member,~IEEE}, Xiaozhe Wang,~\IEEEmembership{Senior Member,~IEEE}}

\maketitle

\begin{abstract}
This paper presents an adaptive stochastic spectral embedding (ASSE) method to solve the probabilistic AC optimal power flow (AC-OPF), a critical aspect of power system operation. The proposed method can efficiently and accurately estimate the probabilistic characteristics of AC-OPF solutions. An adaptive domain partition strategy and expansion coefficient calculation algorithm are integrated to enhance its performance. Numerical studies on a 9-bus system demonstrate that the proposed ASSE method offers accurate and fast evaluations compared to the Monte Carlo simulations. A comparison with a sparse polynomial chaos expansion method, an existing surrogate model, further demonstrates its efficacy in accurately assessing the responses with strongly local behaviors.
\end{abstract}


\section{Introduction}

AC Optimal power flow (AC-OPF) is a critical aspect of power system operations, serving crucial roles in security assessment, unit commitment, economic dispatch, etc. As the rising integration of volatile renewable energy sources (RESs) and new forms of loads (e.g., electric vehicles)  complicates real-time AC-OPF solutions, 
probabilistic AC-OPF has been favored over deterministic ones \cite{Ly2023}. 

To assess uncertainties in the probabilistic AC-OPF problem, Monte Carlo (MC) simulations are widely used due to their straightforward implementation \cite{Zou2014}. However, they can be computationally intensive to achieve high accuracy, rendering them impractical. While analytical methods (e.g., Cumulant \cite{Schellenberg2005,Tamtum2009}) offer improved efficiency, they might face challenges in accurately estimating non-Gaussian-like response distributions.
Deep learning techniques \cite{Pan2023}, \cite{Gao2023} have been developed to efficiently address the AC-OPF problem by mapping uncertainties to generator set points. Although they offer accurate pointwise OPF solutions, they might lack comprehensive interpretability of uncertainty. 


Surrogate models (e.g., Gaussian process regression (GPR) \cite{Pareek2021} and polynomial chaos expansion (PCE) \cite{Mühlpfordt2019,Ly2023}), are favored for delivering accurate AC-OPF solutions with a limited number of sample evaluations and offering clear uncertainty interpretability, e.g., closed-form statistics (mean and variance) representations of OPF solutions. 
For example, Pareek et al.  \cite{Pareek2021} proposed a GPR-based method to solve the probabilistic AC-OPF that reduces the required sample size. 
However, GPR can falter with noisy, multi-modal distributions \cite{Rajabi2019review}.  Mühlpfordt et al. \cite{Mühlpfordt2019} presented a PCE model for AC-OPF, deriving mean and variance directly from PCE coefficients. Yet, the efficiency of the method may decrease as the number of uncertainty inputs grows. 
To address this, Sheng et al. \cite{Sheng2018} and Ly et al. \cite{Ly2023} enhanced PCE methods with adaptive sparse and Lite-PCE schemes, respectively. Still, capturing strongly localized response behaviors, characterized by noisier and more multi-modality, remains challenging, often necessitating higher PCE orders and increasing computational demands.

In this paper, we propose an adaptive stochastic spectral embedding (ASSE) method to accurately and efficiently solve the probabilistic AC-OPF problem under uncertainties arising from RESs and loads. Specially, the ASSE method adaptively establishes a series of spectral expansions, using PCE,  targeting incrementally refined subdomains of the uncertain input space. 
An adaptive domain partition strategy and expansion coefficient calculation algorithm are integrated to enhance its performance.
Simulation results on a 9-bus system demonstrate that the proposed ASSE method offers accurate probabilistic characteristics estimation when compared with MC simulations. Furthermore, ASSE achieves lower validation errors with lower PCE order (e.g., typically 2), outperforming the SPCE method, especially for non-Gaussian-like distributions, and the tail distributions.

The remainder of this paper is organized as follows. Section \ref{sec:AC-OPF} introduces the mathematical formulation of the probabilistic AC-OPF problem. Section \ref{sec:ASSE} elaborates on the ASSE representation to solve the probabilistic AC-OPF. Section \ref{sec:ASSE_ACOPF} presents a detailed ASSE-based approach for the probabilistic AC-OPF. Section \ref{sec: Simulation} shows the numerical studies. Section \ref{sec:cons} gives the conclusions.

%

\section{The Probabilistic AC-OPF } \label{sec:AC-OPF}
\subsection{Probabilistic AC-OPF}
Considering a $N$-bus transmission system, the general AC-OPF problem can be formulated as minimizing the cost of generation while satisfying the physical and operational constraints \cite{Gao2023physics}. Let $\bm{u} = [{P_{G_i}},{Q_{G_i}},V_i], i \in \mathbb{G}$, $P_{G_i}$ and $ Q_{G_i}$ are the active and reactive power output at generator $i$, respectively; $ V_i $ is the corresponding voltage magnitude; $\mathbb{G}$ is the generator index set. 
Let $\bm{\zeta} = [\zeta_1,\cdots,\zeta_{\mathcal{M}}]$ denote the uncertain sources vector, such as wind speed, solar irradiation, or new forms of load (e.g., electrical vehicles).  Due to the existence of randomness, all variables  (e.g.,$\bm{u}$, $ \bm{x}$) become random variables. Then,  the probabilistic AC-OPF is formulated as:
\begin{subequations}
\label{eq:POPF}
\begin{align}
    Y &= \min_{\bm{u}(\bm{\zeta})} \quad  \sum_{i\in \mathbb{G}} C_i (P_{G_i} (\bm{\zeta}))\label{eq:objectives2} \\
    \mathrm{s.t.}&  \quad  {\bm{f}}\left( \bm{x}(\bm{\zeta}), \bm{u}(\bm{\zeta})  \right) =0 ,  \quad  
    \bm{h}(\bm{x}(\bm{\zeta}), \bm{u}(\bm{\zeta})) \leq 0 \label{eq:inequalitys}    
 \end{align} 
 \end{subequations}
 \normalsize
where \eqref{eq:objectives2} is the objective function, with $ C_i(P_{G_i}) = C_{i,2}P_{G_i}^2 + C_{i,1}P_{G_i} + C_{i,0}$ describing the production cost, $P_{G_i}$ denotes the active power output of $i$-th conventional generator; $C_{i,g}, g= \{0,1,2\}$ are the generation cost coefficients for generator $i$. $\bm{x} = [\bm{V},\bm{\theta}]$ denotes the state variables, with $V_i$ and $\theta_i$ being the voltage magnitude and voltage angle, respectively. 
The equality constraint in \eqref{eq:inequalitys} denotes the probabilistic power flow equations. 
%
The inequality constraint in \eqref{eq:inequalitys} includes the voltage limits, branch flow limits, and generation capacity limits considering uncertainties. 
%
Conventional methods for evaluating the stochastic properties of probabilistic AC-OPF solutions typically rely on running a large number of MC simulations, 
which can be computationally inefficient. 
In the following section, an adaptive stochastic spectral embedding method by integrating a spectral expansion (i.e., sparse PCE) scheme is introduced to approximate the AC-OPF model \eqref{eq:POPF}. As such, the AC-OPF problem can be solved more efficiently while maintaining the convergence of global spectral methods (i.e., the PCE method in this paper) and capturing responses with strongly localized behavior. 

\section{Adaptive Stochastic Spectral Embedding} 
\label{sec:ASSE}
This section introduces an ASSE representation to serve as a surrogate model for the probabilistic AC-OPF model. The PCE is employed to approximate the residual expansion of SSE such that the SSE could guarantee the global convergence of the PCE method and maintain the PCE's closed-form representation of statistics\color{black}.  An adaptive sequential partition strategy based on Sobol' index is elaborated for domain partitioning,  based on which, PCE models are constructed in each subdomain. This guarantees the accuracy of SSE in capturing the local behavior of response distributions compared with the SPCE method. \color{black}
Furthermore, to improve the computation efficiency while guaranteeing accuracy, the least angle regression (LAR) together with a hyperbolic truncation scheme is integrated to calculate the PCE coefficients.   
\subsection{The SSE Representation}
Let 
${Z}\color{black} ={F}(\bm{\zeta})$ denote the probabilistic AC-OPF model \eqref{eq:POPF} where ${Z}\color{black}$ could be any of the \color{black}
stochastic responses, i.e., the solutions of \eqref{eq:POPF} (e.g., the decision variables $\bm{u}$, state variables \color{black} $\bm{x}$, 
the objective function $Y$), and $\bm{\zeta}$ denote the random inputs. Assume $Z$ has finite second-order moments (i.e., $\mathbb{E}[{Z}^2] < + \infty$). 
  Let ${\mathcal{D}_{\bm{\zeta}}^{\alpha}}$ be a subdomain of $\mathcal{D}_{\bm{\zeta}}$ (i.e., the entire domain of $\bm{\zeta}$).
Then, ${Z} = {F}(\bm{\zeta})$ can be approximated by a stochastic spectral embedding (SSE) representation through sequentially expanding residuals  $\widehat{\mathcal{R}}_{\mathrm{S}}^{\alpha}(\bm{\zeta})$ in the subdomain ${\mathcal{D}_{\bm{\zeta}}^{\alpha}}$ 
\cite{Wagner2022rare}:
\begin{equation}
\label{eq:SSE}
{Z} = {F}(\bm{\zeta}) \approx 
{F}_{\mathrm{SSE}}(\bm{\zeta})=\sum_{\alpha \in \mathcal{A}} \bm{1}_{\mathcal{D}_{\bm{\zeta}}^{\alpha}}(\bm{\zeta}) \widehat{\mathcal{R}}_{\mathrm{S}}^{\alpha}(\bm{\zeta}) 
\end{equation}
where $\alpha = (k,d)$ with $k = \{0,\cdots,K\}$ denoting the expansion level and $d = \{1,\cdots,D_k\}$ being a subdomain index at a specific level $k$. $K$ is the \color{black} largest number of expansion level and $D_k$ indicates the \color{black} total \color{black} number of subdomains under specific expansion level $k$.
$\bm{1}_{\mathcal{D}_{\bm{\zeta}}^{\alpha}}$ denotes the subdomain index function, and $\bm{1}_{\mathcal{D}_{\bm{\zeta}}^{\alpha}} = 1$ in the $\alpha$-th subdomain and $\bm{1}_{\mathcal{D}_{\bm{\zeta}}^{\alpha}} = 0$ 
 everywhere else. $\widehat{\mathcal{R}}_{\mathrm{S}}^{\alpha}(\bm{\zeta})$ denote the truncated expansions of residuals of the SSE.  To select a refinement domain and suitable error measures, separate the domains into domains that have already been split, and terminal domains that have not been split. Let $\mathcal{J}$ be the terminal domains index set. Then, the  SSE in \eqref{eq:SSE} can be reformulated as:
 \begin{equation}
\label{eq:SSE_sep}
{F}_{\mathrm{SSE}}(\bm{\zeta})=\sum_{\alpha \in \mathcal{A}\backslash{\mathcal{J}}} \bm{1}_{\mathcal{D}_{\bm{\zeta}}^{\alpha}}(\bm{\zeta}) \widehat{\mathcal{R}}_{\mathrm{S}}^{\alpha}(\bm{\zeta}) + \sum_{\alpha \in {\mathcal{J}}} \bm{1}_{\mathcal{D}_{\bm{\zeta}}^{\alpha}}(\bm{\zeta}) \widehat{\mathcal{R}}_{\mathrm{S}}^{\alpha}(\bm{\zeta})
\end{equation}
\normalsize
 
\subsection{The PCE-based Residuals Expansion} \label{sec:PCE}
 The  SSE method involves a local refinement process where a global spectral expansion is gradually divided into progressively smaller subdomains ${\mathcal{D}_{\bm{\zeta}}^{\alpha}}$. In this paper, PCE, a global spectral representation, is employed for the residuals expansion $\widehat{\mathcal{R}}_{\mathrm{S}}^{\alpha}(\bm{\zeta})$ \cite{Xiu2002wiener}:
\begin{equation}
\setlength{\abovedisplayskip}{2pt}
\setlength{\belowdisplayskip}{2pt}
\label{eq:PCE}
\widehat{\mathcal{R}}_{\mathrm{S}}^{\alpha}(\bm{\zeta}) = \widehat{\mathcal{R}}_{\mathrm{PCE}}^{\alpha}(\bm{\zeta}) = \sum_{l=1}^{L} c_{l}^{{\alpha}}\Phi_{l}^{{\alpha}}(\bm{\zeta})
\end{equation}
where $c_{l}^{{\alpha}}$ indicate the $L$ number of PCE coefficients at $d$-th subdomain of $k$-th expansion level with ${\alpha} =(k,d)$.  $\Phi_{l}^{{\alpha}}(\bm{\zeta})$ is the multivariate orthogonal polynomial basis,  \color{black}  which satisfies the following orthogonal condition: \small $\int_{{\mathcal{D}_{\bm{\zeta}}^{\alpha}}} \Phi_{l}^{{\alpha}}(\bm{\zeta}) \Phi_{s}^{{\alpha}}(\bm{\zeta})\Gamma_{\bm{\zeta}}^{\alpha}(\bm{\zeta})  d{\bm{\zeta}} = 0$ \normalsize for ${l} \neq {s}$, where $\Gamma_{\bm{\zeta}}^{\alpha}(\bm{\zeta}) $ is the joint probability density function (PDF) of $\bm{\zeta}$ in the subdomain $\mathcal{D}_{\bm{\zeta}}^{\alpha}$. Specially,  $\Gamma_{\bm{\zeta}}^{\alpha}(\bm{\zeta}) = \bm{1}_{\mathcal{D}_{\bm{\zeta}}^{\alpha}}(\bm{\zeta})\frac{\Gamma_{\bm{\zeta}}(\bm{\zeta})}{{\mathbb{P}}^{\alpha}}$,  with ${\mathbb{P}}^{\alpha} = \int_{\mathcal{D}_{\bm{\zeta}}^{\alpha}}\Gamma_{\bm{\zeta}}(\bm{\zeta})d\bm{\zeta}$ denoting the input probability mass. \color{black} Typically, $\Phi_l(\bm{\zeta})$ can be constructed based on the full tensor product of univariate orthogonal polynomial basis $\phi_j{(\zeta_j)}$. 
\begin{eqnarray}
\label{eq:Poly_tensor}
\Phi_l(\zeta_1,\cdots,\zeta_{\mathcal{M}}) &=& \prod_{j=1}^{\mathcal{{M}}}\phi_j^{(\beta_{j}^{l})}(\zeta_j)
\end{eqnarray}
\normalsize
where $\beta_{j}^{l}$ is the degree of $\phi_j$ for random variable $\zeta_j$ at $l$-th term. $\phi_j$ can be calculated through assumed probability distributions \cite{Xiu2002wiener}, moment-based methods \cite{Wang2021,Wang2022} or  Stieltjes Procedure \cite{Xu2019PPF}.
The accuracy of the PCE method is assessed by the leave-one-out  (LOO) \color{black} cross-validation error $e_{\mathrm{loo}}$, which can be calculated by \cite{Wag2021bay}:
\begin{equation}
\setlength{\abovedisplayskip}{4pt}
\setlength{\belowdisplayskip}{4pt}
\label{eq:loo}
  e_{\mathrm{loo}} = \frac{1}{N_{\mathrm{ref}}}{\sum_{m =1}^{N_{\mathrm{ref}}}\left[ {\color{black}Z^{(m)}\color{black} - \widehat{\mathcal{R}}^{\sim m}_{\mathrm{PCE}}{(\bm{\zeta}^{(m)}})}\right]^2}  
\end{equation}
where $N_{\mathrm{ref}}$ denotes the number of sample points in the subdomains. $Z^{(m)}$ is the $m$-th sample response (e.g., generator power outputs). 
$\widehat{\mathcal{R}}^{\sim m}_{\mathrm{PCE}}$ is the PCE model built using the training sample points 
omitting the $m$-th sample point $\bm{\zeta}^{(m)}$.\color{black}
\subsection{Adaptive Sequential Partition} \label{sec:partition}
\noindent  \textbf{Refinement Domain Selection:} To choose the refinement domain from the unsplit domain sets $\mathcal{J}$, 
%
an $e_{\mathrm{loo}}$ error-based refinement score $\rho^k$ is first designed: 
 \begin{equation}
 \setlength{\abovedisplayskip}{4pt}
\setlength{\belowdisplayskip}{4pt}
\label{eq:refine_score}
\rho^{k+1,s} = \begin{cases} 
e_{\mathrm{loo}}^{k+1, s} {\mathbb{P}}^{k+1, s}, & \mathrm{ if } \quad  \exists \widehat{{R}}_S^{k+1, s}, \\ e_{\mathrm{loo}}^{k, s} {\mathbb{P}}^{k+1, s},
&\mathrm{ otherwise}\end{cases}
\end{equation}
where 
${\mathbb{P}}^{k+1, s}$ denotes the input probability mass, based on which, the subdomain size $N_{\mathrm{ref}}$ is incorporated \color{black} (e.g., a fixed minimum number of points in the subdomain is set $N_{\mathrm{ref,min}} = 10$ in this paper, i.e., $N_{\mathrm{ref}} \geq 10$). $\widehat{{R}}_S^{k+1, s}$ exists when $N_{\mathrm{ref}} \geq N_{\mathrm{ref,min}}$\color{black}. 
\color{black} Based on \eqref{eq:refine_score}, the refinement domain is selected with the largest $\rho^{k}$, i.e.,
\begin{equation}
\setlength{\abovedisplayskip}{2pt}
\setlength{\belowdisplayskip}{2pt}
\label{eq:refine}
  \rho^{\mathrm{refine}} = \arg \max_{k\in\mathcal{J}}\rho^{k}  
\end{equation} 

\noindent  \textbf{Sobol’ Index-Based Partitioning Strategy:} In this paper, we employ a partitioning approach that selects a split direction to maximize the variability of $\widehat{\mathcal{R}}_{\mathrm{S}}^{\alpha}(\bm{\zeta})$, i.e., maximize the difference in residual empirical variance across the resulting subdomains.  As such, the refinement subdomain ${\mathcal{D}_{\bm{\zeta}}^{\alpha}}$ is bifurcated into two subdomains (i.e., $N_{\mathrm{split}}=2$) with the same distributed probability mass in the direction characterized by the peak first-order Sobol' index $S_{j}$, as given by:
\begin{equation}
\setlength{\abovedisplayskip}{4pt}
\setlength{\belowdisplayskip}{4pt}
\label{eq:Sobol_based}
\sigma^{\color{black}\alpha \color{black}}=\underset{j \in\{1, \cdots, \mathcal{M}\}}{\arg \max } S_j^{\color{black}\alpha \color{black}}  
\end{equation}
A notable benefit of utilizing the PCE approach for residuals approximation is its efficiency in computation. Specifically, when random inputs $\bm{\zeta}$ are mutually independent, $S_{j}$ can be directly derived from the coefficients of the PCE model \cite{Sudret2008global}:
\begin{equation}
\label{eq:FirstOrderSI}
S_{j}=\frac{\mathrm{Var}[\widehat{\mathcal{R}}_{\mathrm{PCE}}({\zeta_j}) ]}{\mathrm{Var}[\widehat{\mathcal{R}}_{\mathrm{PCE}}(\bm{\zeta}) ]}
\approx
\frac{\sum_{l\in \mathbb{I}_{j}} c_l^2}{\sum_{l=2}^L c_l^2}
\end{equation}
where $ \mathbb{I}_{j} = \{ l\in \{ 1,...,L\}: \beta_{j}^{l}\not= 0\}$. The detailed sequential partition algorithms can be found in \cite{Wagner2022rare} (Section 2.2).

\subsection{The Adaptive Sparse Procedure for Calculating $c_l$} \label{sec:PCECoefficients}

\noindent \textbf{\textbf{ Hyperbolic Truncation Strategy:}} Typically, $\Phi_l(\zeta_1,\cdots,\zeta_{\mathcal{M}})$ is constructed by the full tensor product of $\phi_j(\zeta_j)$, using a standard truncation scheme given by: $\sum_{j=1}^{\mathcal{M}}\beta_{j}^{l}\leq H, l = \{1,\cdots, L\}$,  where $H$ is the PCE order. Consequently, the total number of terms $L$ in PCE escalates exponentially with both $H$ and $\mathcal{M}$ according to \small$L=(\mathcal{M}+H)!/(\mathcal{M}!H!)$\normalsize. To this end, we adopt a hyperbolic truncation strategy to decrease $L$ to keep the interactions of lower-order univariate polynomial bases \cite{UQdoc_20_104}: \small $\left(\sum_{j=1}^{\mathcal{{M}}}\left(\beta_{j}^{l}\right)^q\right)^{\frac{1}{q}} \leq H$\normalsize, with $ q\in(0,1)$. $H$ and $q$ can be adaptively chosen according to the stopping criteria (i.e., the modified leave-one-out cross-validation error $e_{\mathrm{cloo}}$ in (23)-(24) \cite{Wang2021}).
%


\noindent \textbf{Coefficients $c_l$ Calculation Algorithm:}  This paper adopts the hybrid least angle regression (LAR) to adaptively determine the PCE coefficients $c_l$. Particularly, LAR together with the hyperbolic truncation strategy is used to identify optimal bases $\Phi_l$.  
After pinpointing the optimal bases, the PCE coefficients $c_l$ are computed using ordinary least squares (OLS). 
See \cite{UQdoc_20_104} for the detailed procedures of Hybrid LAR.


\section{ASSE-based Probabilistic AC-OPF} \label{sec:ASSE_ACOPF}
This section presents a step-by-step description of the proposed algorithm for solving AC-OPF problem. 

\noindent\textbf{Step 1.} 
Input network data. Generate $N_{\mathrm{ED}}$ realizations of random inputs $\bm{\zeta}_{\mathrm{ED}} = (\zeta^{(1)},\cdots,\zeta^{(N_{\mathrm{ED}})})$ ( e.g., wind speeds, solar radiations, and real load power) by the Quasi-Monte Carlo  (QMC) method. \\
\noindent\textbf{Step 2.} 
Solve the AC-OPF problem in \eqref{eq:POPF} to obtain solutions $\bm{Z}_{\mathrm{ED}} = (Z^{(1)},\cdots,Z^{(N_{\mathrm{ED}})})$ through deterministic tools (e.g., Matpower \cite{Zimmerman2011}). Pass the data set $[\bm{\zeta}_{\mathrm{ED}},\bm{Z}_{\mathrm{ED}}]$ to \textbf{Step 3.} \\
\noindent\textbf{Step 3.} Construct the ASSE model in \eqref{eq:SSE}: 
\begin{enumerate}
    \item[3a)] {Initialization}: Let $k=0$, $d=1$, \color{black} $N_{\mathrm{ref,min}} = 10$, $K = 1000$\color{black}. Calculate the initial expansion approximated by the PCE model in \eqref{eq:PCE}. 
    \item [3b)] \color{black} Select the refinement domain via \eqref{eq:refine_score}-\eqref{eq:refine} \color{black} and split the domain using the Sobol' index-based partitioning strategy \color{black} \eqref{eq:Sobol_based}-\eqref{eq:FirstOrderSI} \color{black} (i.e., the algorithm described in Section \ref{sec:partition}).
    \item[3c)] For every subspace:\\
    i) Construct the residual expansion based on the PCE model \eqref{eq:PCE} according to Section \ref{sec:PCE} and the algorithm in Section \ref{sec:PCECoefficients};\\
    ii) Calculate and store the refinement score \eqref{eq:refine_score}. If \color{black} $N_{\mathrm{ref}} \geq N_{\mathrm{ref,min}}$ and $k < K$, $k \leftarrow k+1$ go to \textbf{Step 3b)} and continue the iteration. Otherwise, go to \textbf{Step 4.} \color{black}  
\end{enumerate}
\textbf{Step 4.} Generate a large number of $N_{\mathrm{Val}}$ random input samples ($N_{\mathrm{Val}} \gg N_{\mathrm{ED}}$) and evaluate the probabilistic information (e.g., mean, variance, and PDF, and cumulative distribution function (CDF)) of the AC-OPF problem solutions using the constructed SSE model \eqref{eq:SSE}. 
Calculate the validation error according to:
\begin{equation}
e_{\mathrm{Val}}=\frac{N_{\mathrm{Val}-1}}{N_{\mathrm{Val}}}\left[\frac{\sum_{m=1}^{N_{\mathrm{Val}}}\left(Z^{(m)}-F_{\mathrm{SSE}}(\bm{\zeta}^{(m)})\right)^2}{\sum_{m=1}^{N_{\mathrm{Val}}}\left(Z^{(m)}-\hat{\mu}_{Z_{\mathrm{Val}}}\right)^2}\right]
\end{equation}
where $\hat{\mu}_{Z{_\mathrm{Val}}} = \frac{1}{N_{\mathrm{Val}}}\sum_{m=1}^{N_{\mathrm{Val}}}Z^{(m)}$ represents the sample mean of the response, derived from $N_{\mathrm{Val}}$ random input samples.
 
\section{Simulation Studies}\label{sec: Simulation}
In this section, we evaluate the proposed method using a 9-bus, 3-generator system. Specifically,
1 wind farm and 1 solar PV power plant both with 100 MW are connected to bus \{2,3\}, respectively.  Together with 3 stochastic loads at bus \{5,7,9\}, there are 5 random inputs in total. Particularly, 
it is assumed that the wind speed follows the Weibull distribution \color{black} \cite{Karki2006} \color{black} with the shape and scale parameters being $11.153$ and $ 3.289$, respectively; the solar irradiation follows the Beta distribution \color{black} \cite{Salameh1995} \color{black} with the shape and scale parameters being  $1.7$ and $0.74$, respectively; and the random loads follow the Gaussian distributions \color{black}\cite{Billinton2008} \color{black} with 5\% variations.   \color{black}
For demonstration purposes, we only consider independent random inputs in the simulation studies. 
For comparison, results from MC  simulations conducted with MATPOWER \cite{Zimmerman2011} serve as our benchmark. Results based on the sparse PCE (SPCE) method in \cite{Sheng2018} are used for comparison.  
 All simulations were performed using MATLAB  R202022b on a PC equipped with an Intel Core i7-8700 (3.20GHz) with 16GB RAM. Toolbox UQLab is adopted to build the SSE and PCE models \cite{UQdoc_20_118,UQdoc_20_104}. 

\noindent{\textbf{ASSE Model Construction:}} 
First, we generate random input samples starting from $N_{\mathrm{ED}} = 30$. These samples are used to solve the probabilistic AC-OPF problem \eqref{eq:POPF}, yielding the sample response pairs $[\bm{\zeta}_{\mathrm{ED}},\bm{Z}_{\mathrm{ED}}]$ (\textbf{Step 1 - Step 2}). We then use this dataset to build the ASSE and SPCE models \cite{Sheng2018} (\textbf{Step 3}) for all decision variables $\bm{u}$ and the objective function $Y$. Particularly, the LAR algorithm adaptively sets parameters like PCE order $H$ (0 to 6) and norm $q$ (0.5 to 0.8 in 0.05 increments).

\begin{figure}[!ht]
\vspace{-0.1in}
\setlength{\abovecaptionskip}{-0.5cm}
\setlength{\belowcaptionskip}{0cm}
\centering
\includegraphics[width=0.5\textwidth]{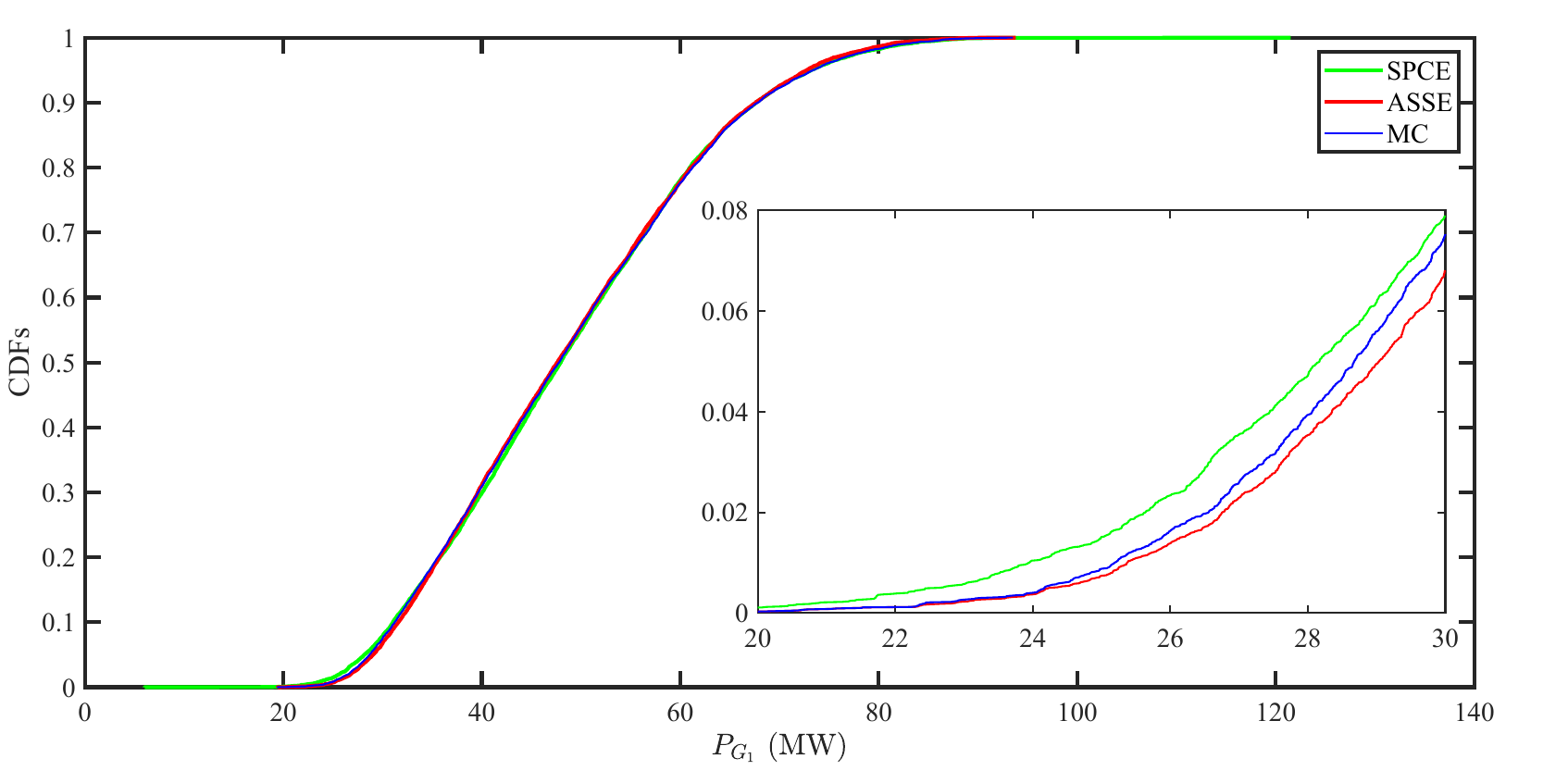}
\caption{The CDFs of $P_{G_1}$  by the ASSE, SPCE, and MC simulations.} 
\label{fig:CDFPDF_Comp_PG1_Case1}
\vspace{-0.1in}
\end{figure}

\begin{figure}[!ht]
\vspace{-0.1in}
\setlength{\abovecaptionskip}{-0.5cm}
\setlength{\belowcaptionskip}{0cm}
\centering
\includegraphics[width=0.51\textwidth]{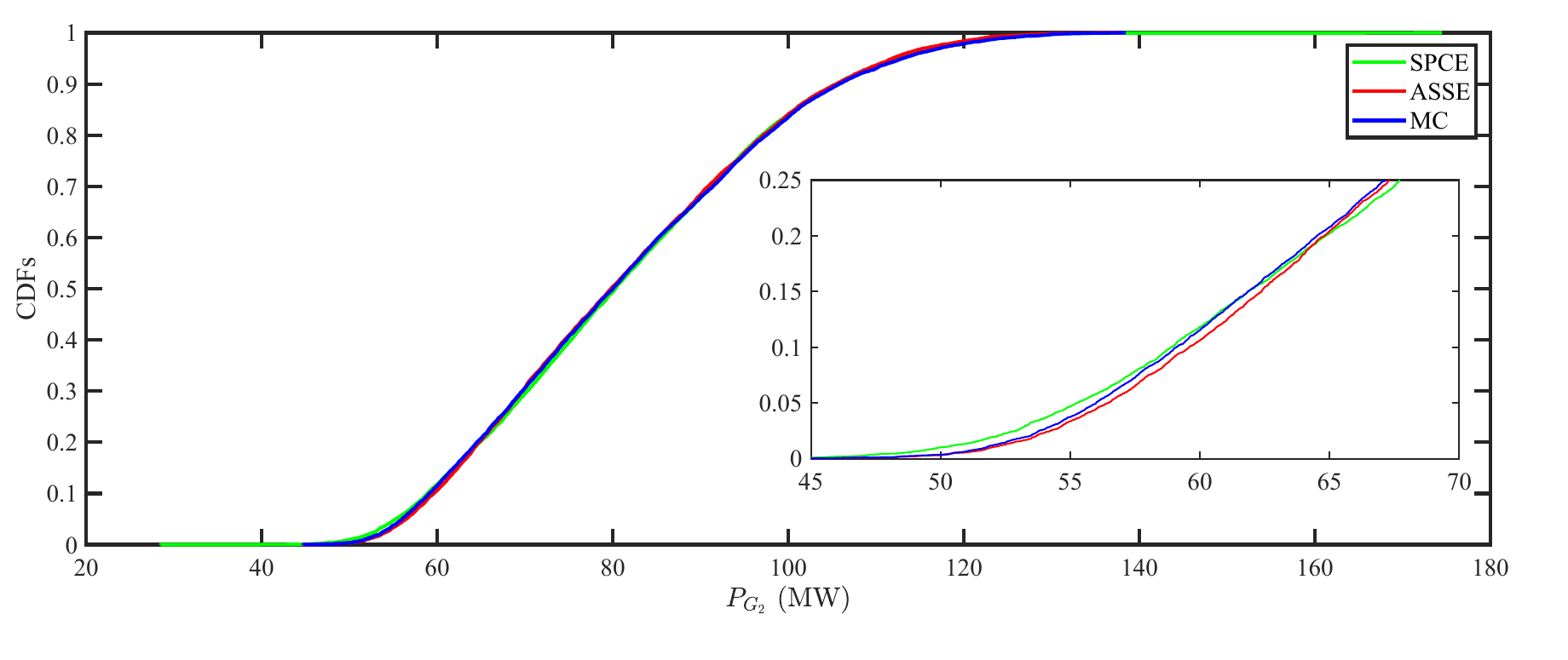}
\caption{The CDFs of $P_{G_2}$  by the ASSE, SPCE, and MC simulations.} 
\label{fig:CDFPDF_Comp_PG2_Case1}
\vspace{-0.1in}
\end{figure}

\begin{figure}[!ht]
\vspace{-0.1in}
\setlength{\abovecaptionskip}{-0.52cm}
\setlength{\belowcaptionskip}{-0.2cm}
\centering
\includegraphics[width=0.5\textwidth]{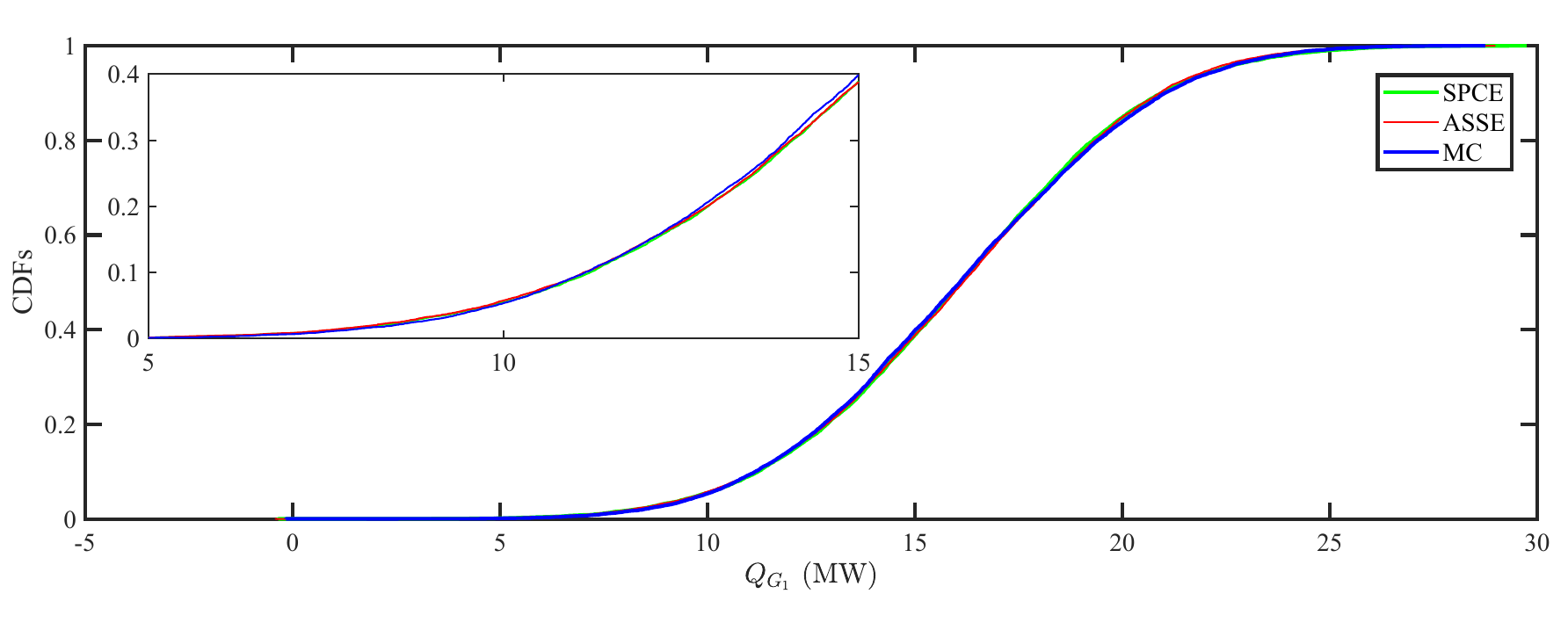}
\caption{The CDFs of $Q_{G_1}$ by the ASSE, SPCE, and MC simulations.} 
\label{fig:CDFPDF_Comp_QG1_Case1}
\vspace{-0.1in}
\end{figure}

\begin{figure}[!ht]
\vspace{-0.1in}
\setlength{\abovecaptionskip}{-0.52cm}
\setlength{\belowcaptionskip}{-0.2cm}
\centering
\includegraphics[width=0.5\textwidth]{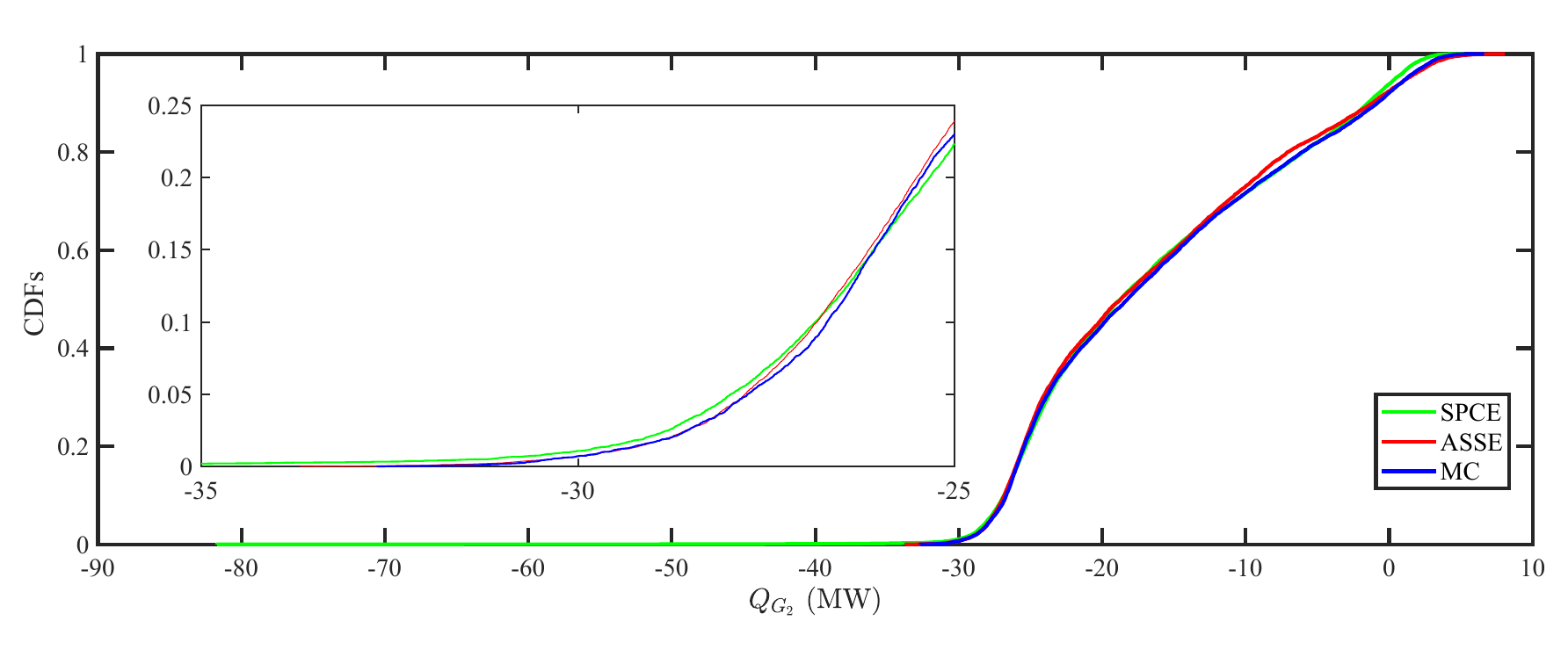}
\caption{The CDFs of $Q_{G_2}$ by the ASSE, SPCE, and MC simulations.} 
\label{fig:CDFPDF_Comp_QG2_Case1}
\vspace{-0.1in}
\end{figure}

\noindent \textbf{Performance Evaluation:} For validation, we generate $N_{\mathrm{Val}} = 10,000$ QMC 
samples of $\bm{\zeta}$ to validate the proposed method. Fig. \ref{fig:CDFPDF_Comp_PG1_Case1}-\ref{fig:CDFPDF_Comp_QG2_Case1} presents the comparison of CDFs of generator power output (i.e., $P_{G_i}$, $Q_{G_i}, i = \{1,2\}$)  from the proposed SSE method (red), the SPCE (green)  \cite{Sheng2018} method, and the benchmark MC simulations (blue). The ASSE results closely align with the MC simulations, as highlighted in 
Fig. \ref{fig:CDFPDF_Comp_PG1_Case1}- \ref{fig:CDFPDF_Comp_QG2_Case1}. Specifically, in Fig. \ref{fig:CDFPDF_Comp_PG1_Case1}-\ref{fig:CDFPDF_Comp_PG2_Case1}, the ASSE employs an order $H =2$, while the SPCE-based model uses $H=3$. For distributions resembling Gaussian, as seen in Fig. \ref{fig:CDFPDF_Comp_QG1_Case1}, both ASSE and SPCE methods exhibit comparable accuracy. However, for responses that exhibit higher localized behavior, the SPCE may require a higher order compared with the ASSE method (e.g., see Fig. \ref{fig:CDFPDF_Comp_QG2_Case1},  $H=2$ for the ASSE-based model, while 
$H=6$ for the PCE-based model, 
and for both models $N_{\mathrm{ED}}= 60$\color{black}). Moreover, the ASSE shows better estimations in terms of the distributions in the tails.

\noindent \textbf{Decisions under Uncertainties:} Table \ref{tab:mean_comp_Dif} presents the estimated mean values of $P_{G_i}$ and the corresponding mean cost value $\mu_Y$ and its normalized error\% ($e_{Y}\%$) from the ASSE, the SPCE, and the MC simulations. These results clearly demonstrate that the mean values and costs estimated from the ASSE are close to those obtained from the MC simulations, and are comparable to those from the SPCE.  Additionally, by analyzing the estimated CDFs of $P_{G_i}$ (as shown in Fig. \ref{fig:CDFPDF_Comp_PG1_Case1}-\ref{fig:CDFPDF_Comp_PG2_Case1}), we can determine $P_{G_i}$ values at 5\% and 95\% confidence intervals, along with their corresponding cost values, as shown in  Table \ref{tab:lower_expectation_comp_Dif}-\ref{tab:upper_expectation_comp_Dif}. Tables \ref{tab:lower_expectation_comp_Dif} to \ref{tab:upper_expectation_comp_Dif} reveal that the ASSE provides closer 
lower and upper bounds to those obtained by the benchmark MC simulations when compared with the SPCE method. 
\color{black}This further validates 
the ASSE's 
slightly superior performance in capturing tail distributions compared to the SPCE. \color{black}

\color{black}
%
%
%
 \begin{table}[!ht]
 \setlength{\abovecaptionskip}{0.cm}
 \setlength{\belowcaptionskip}{-1cm}
 \vspace{-0.3cm}
 \renewcommand{\arraystretch}{1.0}
 \caption{Comparison of the estimated statistics of $P_{G_i}$ and the cost $Y$ by the MC simulations, the ASSE, and SPCE methods}
 \label{tab:mean_comp_Dif}
 \centering
 \resizebox{.96\columnwidth}{!}{
 \begin{tabular}{c|c|c|c|c|c}
 \hline
 Index & $\mu_{P_{G_1}}$ & $\mu_{P_{G_2}}$ & $\mu_{P_{G_3}}$ & $\mu_{Y}$ & $e_{\mu_Y}\%$ \\
 \hline
 MC  & $48.8203$ & 81.5131 & 57.1871  & $2.7117\times 10^{3}$ & -- \\
\hline 
 ASSE   & 48.8583  & 81.5490 & 57.2138 & $2.7132\times 10^{3}$ &-0.0674\\ 
 \hline 
SPCE \cite{Sheng2018} &48.6991  &81.3435 & 57.0702 & $2.7055\times 10^{3}$ & 0.0167 \\ 
 \hline
\end{tabular}}
\vspace{-0.15in}
\end{table}
%
\begin{table}[!ht]
\setlength{\abovecaptionskip}{0.cm}
\setlength{\belowcaptionskip}{-0.5cm}
\renewcommand{\arraystretch}{1.0}
\caption{Comparison of the estimated statistics of the $P_{G_i}$ and the cost $Y$ and its normalized error at 5\% confidence interval (lower bound) by the MC simulations, the ASSE, and SPCE methods}
\label{tab:lower_expectation_comp_Dif}
\centering
\resizebox{.96\columnwidth}{!}{
\begin{tabular}{c|c|c|c|c|c}
\hline
Index & $P_{{G_1}_{{\mathrm{low}}}}$ &$P_{{G_2}_{{\mathrm{low}}}}$ & $P_{{G_3}_{{\mathrm{low}}}}$ & $Y_{\mathrm{low}}$&$e_{Y_\mathrm{low}}\%$ \\
\hline
MC  & $28.6788$ & $56.0014$ &   $39.5553$ & $1.8839 \times 10^{3}$ & -- \\
\hline 
 ASSE   & $28.7079$  &  $56.0314$ &  $39.5743$ & $1.8847 \times 10^{3}$ & 0.0469 \\ 
 \hline 
SPCE \cite{Sheng2018} & $28.8302$ & $56.1134$ & $39.6290 $&$1.8876 \times 10^{3}$& 0.1968 \\ 
 \hline
\end{tabular}}
\vspace{-0.5cm}
\end{table}
\begin{table}[!ht]
\setlength{\abovecaptionskip}{0.cm}
\renewcommand{\arraystretch}{1.0}
\caption{Comparison of the estimated statistics of the $P_{G_i}$ and the cost $Y$ and its normalized error at 95\% confidence interval (upper bound) by the MC simulations, the ASSE, and SPCE methods}
\label{tab:upper_expectation_comp_Dif}
\centering
\resizebox{.96\columnwidth}{!}{
\begin{tabular}{c|c|c|c|c|c}
\hline
Index & $P_{{G_1}_{{\mathrm{up}}}}$ &$P_{{G_2}_{{\mathrm{up}}}}$ & $P_{{G_3}_{{\mathrm{up}}}}$ & $Y_{\mathrm{up}}$&$e_{Y_\mathrm{up}}\%$ \\
\hline
MC  & $73.4004$ & $113.0811$ &   $79.0015$ & $4.1108 \times 10^{3}$ & -- \\
\hline 
 ASSE   & $73.4251$  &  $113.0181$ &  $78.9429$ & $4.1089 \times 10^{3}$ & -0.0476 \\ 
 \hline 
SPCE \cite{Sheng2018} & $73.3637$ & $112.8227$ & $78.7551 $&$4.0998 \times 10^{3}$& -0.2690 \\ 
 \hline
\end{tabular}}
\end{table}
\color{black}
\noindent \textbf{Validation Error Comparison:} Next,
\color{black}
we present the validation error $e_{\mathrm{Val}}$ under different 
training sample size 
$N_{\mathrm{ED}}$ (ranging from 30 to 240 in increments of 5) in estimating $P_{G_1}$ and $Q_{G_2}$. \color{black}  Fig. \ref{fig:ValError_Comp_PG1_Case1} clearly shows that ASSE has a much smaller \color{black} $e_{\mathrm{Val}}$ \color{black}
compared to the SPCE in all different $N_{\mathrm{ED}}$. Furthermore, both methods show very high accuracy when $N_{\mathrm{ED}}$ is large (e.g., $\geq 100$ in this scenario).
\color{black} In Fig. \ref{fig:ValError_Comp_PG1_Case1} (b), when $N_{\mathrm{ED}} = 55, H=4$, the SPCE model overfits, resulting in a sharp increase in $e_{\mathrm{Val}}$. \color{black}
\color{black}

\noindent \textbf{Time Consumption:}  
The proposed ASSE requires much less computational time compared to the MC simulations since  $N_{\mathrm{ED}} \ll N_{\mathrm{Val}} $, e.g., the computational time for the ASSE is about $54.75$s with $N_{\mathrm{ED}} = 60$ while the time for MC simulations is about $9125$s.
 %
%
\begin{figure}[!ht]
\setlength{\abovecaptionskip}{-0.61cm}
\setlength{\belowcaptionskip}{-0.7cm}
\centering
\includegraphics[width=0.50\textwidth]{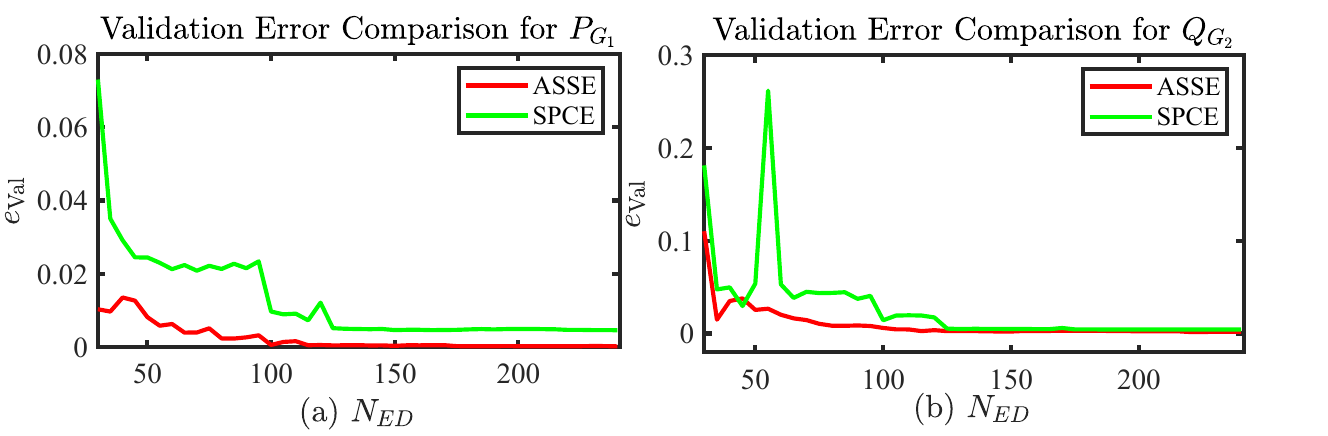}
\caption{Validation errors $e_{\mathrm{Val}}$  of different training sample sizes $N_{\mathrm{ED}}$ for $P_{G_1}$ \color{black} and $Q_{G_2}$ \color{black} by the ASSE, and SPCE method. Left: $P_{G_1}$, Right: $Q_{G_2}$.  } 
\label{fig:ValError_Comp_PG1_Case1}
\vspace{-0.14in}
\end{figure}
\color{black}
\section{Conclusions} \label{sec:cons}
This paper proposed an ASSE method to address the probabilistic AC-OPF problem in the presence of uncertainties. Simulation results demonstrated that the proposed method can provide accurate probabilistic characteristic estimations of OPF solutions (e.g., generator power outputs and the objective function). The efficacy of the proposed method was validated by comparing them with the MC simulations. Furthermore, ASSE effectively captured distributions with higher noise and multi-modality, while ensuring the convergence of the PCE method.
%


\balance

\end{document}